\documentclass{INTERSPEECH2023}

\interspeechcameraready

\usepackage{amsmath}
\usepackage{graphicx}
\usepackage{booktabs}
\usepackage{array}

\usepackage{xcolor}
\usepackage{cleveref}
\usepackage{cite}
\usepackage{bm}
\usepackage[ruled,vlined]{algorithm2e}
\usepackage{multirow}
\usepackage{indentfirst}

\graphicspath{./figures}

\title{Comparing normalizing flows and diffusion models for prosody and acoustic modelling in text-to-speech}

\name{
Guangyan Zhang$^{1,2}$, Thomas Merritt$^{1}$, Manuel Sam Ribeiro$^{1}$, Biel Tura-Vecino$^1$, \\Kayoko Yanagisawa$^1$,
Kamil Pokora$^1$, Abdelhamid Ezzerg$^1$, Sebastian Cygert$^{1,3}$, Ammar Abbas$^1$,
\\Piotr Bilinski$^{1,4}$, Roberto Barra-Chicote$^1$, Daniel Korzekwa$^1$, Jaime Lorenzo-Trueba$^1$
\thanks{Work conducted when all authors were at Amazon TTS Research.}
}

\address{
  ~$^1$ Amazon TTS, ~$^2$ Department of Electronic Engineering, The Chinese University of Hong Kong,\\ ~$^3$ Gdańsk University of Technology, ~$^4$ University of Warsaw
  }

\email{\{manuerib, truebaj\}@amazon.com}

\begin{document}

\maketitle
\begin{abstract}
Neural text-to-speech systems are often optimized on $\mathcal{L}$1/$\mathcal{L}$2 losses, which make strong assumptions about the distributions of the target data space.
Aiming to improve those assumptions, Normalizing Flows and Diffusion Probabilistic Models were recently proposed as alternatives. In this paper, we compare traditional $\mathcal{L}$1/$\mathcal{L}$2-based approaches to diffusion and flow-based approaches for the tasks of prosody and mel-spectrogram prediction for text-to-speech synthesis. We use a prosody model to generate \emph{log-f0} and duration features, which are used to condition an acoustic model that generates mel-spectrograms. Experimental results demonstrate that the flow-based model achieves the best performance for spectrogram prediction, improving over equivalent diffusion and $\mathcal{L}1$ models. Meanwhile, both diffusion and flow-based prosody predictors result in significant improvements over a typical $\mathcal{L}2$-trained prosody models.
\end{abstract}

\section{Introduction}
\label{sec:intro}

Neural text-to-speech (TTS) has recently demonstrated significant success in generating high-quality and stable speech \cite{elias2021parallel,tan2021survey,shah2021non,ren2020fastspeech,lancucki2021fastpitch}. However, TTS systems are still affected by the one-to-many mapping problem caused by speech containing many possible variations not directly explained by the phoneme sequence, such as prosody\cite{tan2021survey} or emotion\cite{zhang2021estimating}.
The typical approach of training with $\mathcal{L}1$ or $\mathcal{L}2$ losses pushes the model to produce `average' (over-smoothed) mel-spectrograms, resulting in synthesised speech with flat prosody and low quality \cite{sheng2019reducing}. Two strategies can be applied to handle this problem. On the one hand, we may provide auxiliary inputs to the acoustic model, such as explicit prosodic features \cite{ren2020fastspeech,lancucki2021fastpitch}. This has the additional advantage of allowing disentangled prosody control, or transfer\cite{zhang2022iemotts}. Alternatively, we may move from a point-based estimation approach (e.g. models using $\mathcal{L}1$ or $\mathcal{L}2$ losses) to a probability density estimation approach (models using Normalizing Flows \cite{miao2020flow, kim2020glow} or Diffusion Models \cite{popov2021grad, jeong2021diff}).
We compare both strategies and analyze their impact on text-to-speech synthesis.

In terms of the first strategy, we may use two acoustic correlates of prosody, duration and $f0$, to explain speech variation.
This helps the model to disambiguate the one-to-many problem of TTS synthesis. 
Oracle prosodic features extracted from the target speech can be provided to the model during training. However, for inference a prosody predictor is required to provide the prosodic features to the acoustic model. In previous work \cite{ren2020fastspeech, lancucki2021fastpitch, kenter2019chive}, the $\mathcal{L}2$ loss function is applied to optimize the prosody predictors. However, this loss results in the predictor generating average prosody, which lacks expressivity.
In \cite{hodari2019using}, a %
Mixture Density Network (MDN), is used to overcome the prosody prediction over-smoothing problem.
However, the authors investigate only $f0$, and, for acoustic modelling, rely on an older Statistical Parametric Speech Synthesis (SPSS) system, rather than a more recent neural TTS approach.
More recently, \cite{abbas2022expressive} found encouraging results when applying normalizing flows to the task of duration modelling. However, this prediction was independent of $f0$. In this paper, state-of-art generative models, normalizing flow and diffusion, are investigated for the task of joint $f0$ and duration modelling.

It is the \emph{goal of this paper} to compare a traditional $\mathcal{L}1$/$\mathcal{L}2$-loss approach to probability density estimation approaches using Normalizing Flows and Diffusion Models.
We investigate these architectures for the tasks of prosody modelling (generation of \emph{f0} and duration) and acoustic modelling (generation of mel-spectrograms).
Our key contributions are:
1) the use normalizing flows and diffusion models to address the problem of over-smoothing for joint $f0$ and duration prediction; and
2) a direct comparison of typical $\mathcal{L}1$/$\mathcal{L}2$-loss based approaches to normalising flows and diffusion models for the tasks of acoustic and prosody modelling. 

\begin{figure*}[t]
  \centering
  \includegraphics[width=0.7\textwidth]{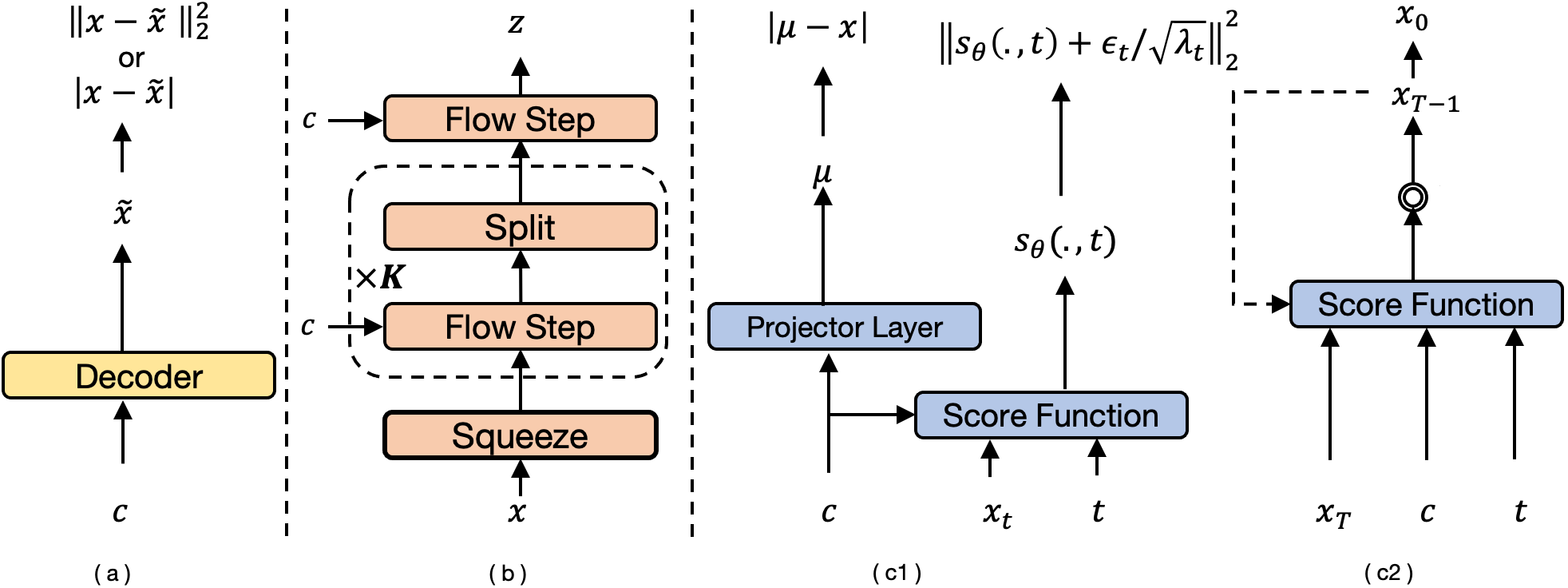}
  \caption{(a)  $\bm{\mathcal{L}1}$ \textbf{or} $\bm{\mathcal{L}2}$ \textbf{Loss}\textbf{-based Model} (b) \textbf{Normalizing Flow Model} (c) Training and inference stages of \textbf{Diffusion Model}}
  \label{fig:various_models}
\end{figure*}

\section{Models}

\label{various_models}
We investigate normalizing flows and diffusion models for the tasks of prosody and acoustic modelling. %
These are compared against standard $\mathcal{L}1$/$\mathcal{L}2$-based approaches from literature.

 $\bm{\mathcal{L}1}$\textbf{/}$\bm{\mathcal{L}2}$ \textbf{loss} is commonly applied in literature during model training. %
\autoref{fig:various_models}(a) illustrates how conditional information $\mathbf{c}$ is converted to the predicted target feature $\mathbf{\Tilde{x}}$ by a non-autoregressive decoder. The structure of the decoder is described in~\cite{shah2021non}. The model is optimized by minimizing the $\mathcal{L}1$ (for spectrogram prediction) or $\mathcal{L}2$ (for prosody prediction) loss between $\mathbf{\Tilde{x}}$ and the target feature $\mathbf{x}$. 
From a probabilistic perspective, minimizing $\mathcal{L}1$ and $\mathcal{L}2$ loss is equivalent to maximizing log-likelihood with Laplacian and Normal distributions.
This strong assumption often results in an ``over-smoothed" prediction, e.g., flat speech or a blurred image. This loss is typically applied to train both prosody prediction (i.e. duration and $f0$) and spectrogram prediction models.

\textbf{Normalizing Flows} have demonstrated state-of-the-art performance for TTS\cite{miao2020flow,kim2020glow,bilinski2022creating} and voice conversion\cite{merritt2022text, bilinski2022creating}.
For this study, we use the Flow-TTS model topology first described in \cite{miao2020flow}, illustrated in \autoref{fig:various_models}(b). 
Pre-trained phoneme alignments are used instead of using attention following \cite{merritt2022text,bilinski2022creating,ezzerg2023remap}. 
During training, the flow model learns a transformation of the target feature $\mathbf{x}$ into the latent variable $\mathbf{z}$ using a series of invertible flow steps $f^{-1}$. 
Conditioning features, $\mathbf{c}$, are provided at each of the flow steps using affine coupling blocks. 
The flow is trained to maximise the likelihood that $z$ comes from a prior distribution. This means the flow learns to map from an unknown complex distribution of $x$, to a vector which comes from a simpler prior distribution. 
This allows the model to optimize for the exact log-likelihood of the data distribution. 
This is in contrast to $\mathcal{L}1$/$\mathcal{L}2$ losses which place a strong assumption on the distribution of $x$ directly. 
For this investigation we use a simple $\mathcal{N}(0,1)$ prior distribution. 
During inference, the predicted features $\mathbf{\Tilde{x}}$ can be derived by sampling from the prior distribution.

\textbf{Diffusion models} \cite{song2020score,popov2021grad}, similarly to flows, %
learn to convert from a simple prior distribution to the unknown complex target feature distribution corresponding to the conditional information. Here we omit the mathematical derivation of the diffusion models and refer the reader instead to \cite{song2020score,popov2021grad} for further information. During training, noise sampled from the distribution $\mathcal{N}(\boldsymbol{\mu}, \boldsymbol{I})$ is repeatedly added to the target feature $\boldsymbol{x}$ at each layer of the diffusion model. %
As a result, eventually $x$, which comes from an unknown complex distribution, is transformed to %
the noise distribution as the number of layers $t \rightarrow \infty $. 
By solving the SDE in \cite{popov2021grad}, %
the vector 
$\boldsymbol{x}_t$, %
output by layer t, 
can be derived directly without intermediate noisy samples $\boldsymbol{x}_1, \dots, \boldsymbol{x}_{t-1}$. %

In parallel, a model is trained to predict the noise that was added at each of the layers of the diffusion network, referred to as the score-based model. %
The predicted noise is subsequently removed from $\boldsymbol{x}_t$, enabling for a mapping from the noise distribution back to the target data $x$. 
During training, the mean of the prior distribution $\boldsymbol{\mu}$ is %
predicted from the conditional information $\mathbf{c}$ by a projector layer, as shown in \autoref{fig:various_models} (c1). An $\mathcal{L}1$ loss is applied between $\boldsymbol{\mu}$ and $x$, %
meaning that the prior of the noise is an over-smoothed spectrogram representation (similar to that learnt by the $\mathcal{L}1$/$\mathcal{L}2$ approaches described above). 
Note that we also attempted to use an uninformative $\mathcal{N}(0,1)$ Gaussian distribution for the noise prior, however the model performed worse and required a larger number diffusion layers.  The score-based model $s_\theta(\boldsymbol{x}_t, \boldsymbol{c}, t)$ is implemented using a U-Net\cite{ronneberger2015u} architecture with $\boldsymbol{c}$, layer $t$ and noisy sample $\boldsymbol{x}_t$ as the inputs. Thus, the loss function consists of two parts, a weighted score-matching objective corresponding to estimating the score function of $p (\boldsymbol{x}_t)$ and an $\mathcal{L}1$ loss between $\boldsymbol{\mu}$ and target feature $\boldsymbol{x}$. %

\section{Prosody and Acoustic Modelling}

\subsection{Prosody Models}
\label{prosody_predictors}
We investigate the three model architectures described in Section~\ref{various_models} for the task of prosody modelling.
We define the task of prosody prediction as the joint modelling of phoneme-level \emph{log-f0} and duration. Frame-level \emph{log-f0} is linearly interpolated over unvoiced regions and then mean-normalized at the speaker level. Given the forced-aligned phoneme sequences, we average \emph{log-f0} at the phoneme-level.
Duration is measured by the number of frames aligned to each phoneme.

All models share an identical encoder architecture which processes the input conditioning features. We vary the decoders and optimization steps, following Section~\ref{various_models}. The input to the models consists of the phoneme sequence and a categorical speaking style identifier. The encoder for the phoneme sequence follows an identical architecture to the one described in \cite{shah2021non}. 
The one-hot speaking style identifier is transformed by an embedding layer and concatenated to the output of the phoneme encoder.
The encoder produces the embedding $\mathbf{c}$, which is used to condition the decoder of the investigated decoder architectures. %
The prosody models output two-dimensional vectors, corresponding to phoneme-level \emph{log-f0} and duration.
We use phoneme-level \emph{log-f0} modelling, as a preliminary analysis %
found that an acoustic model conditioned with oracle phoneme-level \emph{log-f0} performs slightly better or identically to oracle frame-level \emph{log-f0}.
Additionally, the phoneme-level approach has the advantage of allowing joint prediction of \emph{log-f0} and duration, which can better capture the relationship between the two prosodic features.

\subsection{Acoustic Models}
\label{acoustic_models}

We investigate the three model architectures in Section ~\ref{various_models} also for the task of acoustic modelling. The inputs to all acoustic models are the phoneme sequence, a pre-trained %
speaker embedding\cite{wan2018generalized}, phoneme-level \emph{log-f0} and phoneme durations. 
Unlike the prosody models, the acoustic models are not conditioned with speaking style information. This is because style is largely conveyed by prosody, also speaker and style attributes are highly entangled in the dataset used. 
The acoustic models are optimized on the target mel-spectrograms, extracted from the time-domain waveform. 
As above, the acoustic models share the same encoder, but use differing decoders. 
The encoder uses the same model architecture as the prosody models. %
However, unlike the prosody models, the phoneme-level conditional encoding $\mathbf{c}$ is upsampled to the frame level before being passed to the decoder. The speaker embedding and phoneme-level \emph{log-f0} features are concatenated to the phoneme encodings. 
At training time, oracle phoneme-level \emph{log-f0} and durations are provided to the acoustic model, while at synthesis-time, we use features generated by one of the prosody models.

\section{Experimental Protocol}
Throughout our experiments, we use a internal dataset of 200 speech hours, recorded by 116 native speakers of English, across a variety of expressive speaking styles such as happiness, sadness, anger, etc.
A sampling rate of 24kHz was used for all recordings, from which 80-dimensional mel-spectrograms were extracted with a frame length of 50 ms and a frame shift of 12.5 ms. 
We use a universal neural vocoder \cite{jiao2021universal} to map generated mel-spectrograms to time-domain waveforms.

To simplify the number of comparisons, we first evaluate the three acoustic model architectures conditioned with oracle prosody features.
We then select the best acoustic model and compare the different prosody models.
Following Section~\ref{acoustic_models}, we consider three acoustic models.
1) \textit{$\mathcal{L}1$-AM}:  acoustic model trained with $\mathcal{L}1$ Loss. An $\mathcal{L}1$ loss-based model is investigated instead of  $\mathcal{L}2$ following recent work \cite{ren2020fastspeech, yu2020durian, lancucki2021fastpitch}. 2) \textit{Flow-AM}: Flow-based acoustic model. 3) \textit{Diff-AM}: Diffusion-based acoustic model.
In addition, we define an upper-bound by copy-synthesis, generating time-domain waveforms from oracle mel-spectrograms with the universal neural vocoder. This system is termed \textit{ORA-AM}.
Once the best performing acoustic model is selected, we compare the different prosody models.
1) \textit{$\mathcal{L}2$-PM}: prosody model trained with $\mathcal{L}2$ loss. %
This is selected %
as it features heavily in recent studies \cite{ren2020fastspeech,kingma2018glow,kenter2019chive, shah2021non}.
2) \textit{Flow-PM}: Flow-based prosody model.
3) \textit{Diff-PM}: Diffusion-based prosody model. 
As before, an upper-bounded system \textit{ORA-P} is created by feeding the acoustic model with oracle prosody features.

We conduct subjective evaluations of the models using a MUSHRA standard reference evaluation paradigm, considering \textbf{Naturalness}, \textbf{Style Similarity} and \textbf{Expressiveness}.
Each listening test included 300 utterances generated by each of the competing systems.
Utterances were rated by 300 native speakers using a crowdsourcing platform.
Each listener rated 15 MUSHRA screens.
We test for statistical significance between systems using paired t-tests with Holm-Bonferroni correction applied. 
All reported significant differences are at the level of $p < 0.05$. 
We assign Naturalness and Style Similarity higher importance because we consider higher expressiveness to be favored only if there are no impacts on naturalness and style similarity.
We also adopt objective metrics to further analyze the generated prosody features.
We observe the standard deviation (STD) of \emph{log-f0}, $\Delta$\emph{log-f0} and duration.
These statistics mirror the dynamics of the prosody features that can be associated with the expressiveness of speech\cite{clark1999using}.
Additionally, we apply the Jensen-Shannon divergence (JSD)\cite{abbas2022expressive} %
to measure the distance between oracle and generated features.

\section{Experimental Results}

\subsection{Acoustic Models}

For inference with Flow and Diffusion models, we sample a latent variable from a prior distribution.
The temperature $\tau$, i.e., standard deviation, of that distribution can impact the quality of generated speech\cite{popov2021grad, kim2020glow}.
Typically, high temperature values, such as $\tau=1$,  bias the model to produce more varied speech, but can negatively impact quality. 
Meanwhile, a low temperature value often results in flatter intonation\cite{kim2020glow}. 
We investigate the %
temperature which best manages the trade-off of expressivity and quality 
for the Flow and Diffusion-based models by %
conducting naturalness subjective evaluations. 
We consider  $\tau \in \{0.2, 0.4, 0.6, 0.8\}$ and present results in \autoref{acoustic_temps_comp}.
The highest temperature is chosen when its corresponding Naturalness MUSHRA scores have no statistically significant difference from the highest MUSHRA scores. Therefore, $\tau=0.4$ and $\tau=0.8$ were selected for \textit{Flow-AM} and \textit{Diff-AM}, respectively.
We also observe that $\tau$ has a much larger impact for the Flow-based system than for the Diffusion-based system, implying that careful temperature tuning is especially important for Flow-based systems.

\begin{table}[htbp]
        \caption{MUSHRA naturalness evaluation results for temperature $\tau$, showing mean values with 95\% confidence intervals. * indicates no statistically significant difference from the highest MUSHRA scores.}
    \centering
    \scalebox{0.8}{
    \begin{tabular}{m{1.25cm} m{1.6cm}m{ 1.6cm}  m{1.6cm}  m{1.6cm}}
         \toprule 
         System & $\tau=0.2$ & $\tau=0.4$ & $\tau=0.6$ & $\tau=0.8$\\
         \midrule
         \textit{Flow-AM} & $\mathbf{77.95\pm1.12}$ & ${77.39\pm1.13}^{*}$ & ${77.03\pm1.14}$ & ${72.86\pm1.41}$\\
         \textit{Diff-AM} & ${78.13\pm1.12}$ & ${79.20\pm1.01}$ & ${79.79\pm1.06}^{*}$ & $\mathbf{80.01\pm1.05}$\\
         \bottomrule
    \end{tabular}
    }
    \label{acoustic_temps_comp}
\end{table}

Using the selected temperatures for the Flow- and Diffusion-based systems, we %
compare all acoustic models conditioned with oracle \emph{f0} and duration. %
Results for all systems across the three evaluation metrics are presented in \autoref{acoustic_comp}. 
In terms of naturalness, there are no significant differences between the three acoustic models.
\textit{ORA-AM} is significantly preferred over \textit{$\mathcal{L}1$-AM} and \textit{Diff-AM}. 
However, there is no significant preference between \textit{ORA-AM} and \textit{Flow-AM}. 
In terms of style similarity, there is no significant difference between \textit{Flow-AM} and \textit{$\mathcal{L}1$-AM}, however both outperform \textit{Diff-AM}. In terms of expressiveness, \textit{Diff-AM} is found to significantly outperform %
the remaining two systems, while there is no significant difference between \textit{Flow-AM} and \textit{$\mathcal{L}1$-AM}. We hypothesize that \textit{ORA-AM} is rated higher than all three systems in terms of style similarity and expressiveness because, in addition to the oracle prosody features, some speaking styles are expressed by alternative acoustic attributes (e.g., laughing). It is somewhat surprising that \textit{Diff-AM} achieves higher expressiveness but lower style similarity scores than the other two systems. A possible explanation for this is that we are using a higher temperature for the Diffusion-based acoustic model, which may come at the cost of style similarity.

To investigate our results further, we conduct a naturalness preference test on the two best performing systems: \textit{$\mathcal{L}1$-AM} and \textit{Flow-AM}.
Relative preference scores for \textit{$\mathcal{L}1$-AM}, \textit{Flow-AM}, and \textit{No Preference} are $24.72\%$, $29.33\%$, and $45.95\%$ respectively.
A binomial significance test with the \textit{No Preference} scores divided equally amongst the two competing systems indicates a statistically significant preference for \textit{Flow-AM} at the level of $p < 0.05$. 
Therefore, %
the Normalizing Flow system 
is found to provide the best results overall.
Consequently, \textit{Flow-AM} is %
selected to evaluate the prosody models. %

\begin{table}[htbp]
        \caption{Mean MUSHRA scores for acoustic models using oracle \textit{f0} and duration, with 95\% confidence intervals.}
    \centering
    \scalebox{0.8}{
    \begin{tabular}{cccc}
         \toprule 
         Method             &  Naturalness              & Style Similarity          & Expressiveness \\
         \midrule
         \textit{ORA-AM}       & $78.2\pm1.11$             & $79.75\pm0.97	$            & $83.16\pm0.83$ \\
         \midrule
         \textit{$\mathcal{L}1$-AM}    & $76.87\pm1.13$            & $77.17\pm1.07$            & $78.65	\pm0.96	$ \\
         \textit{Flow-AM}   & $\mathbf{77.07\pm1.11}$   & $\mathbf{77.29\pm1.04	}$   & $78.69	\pm0.97	$ \\
         \textit{Diff-AM}   & $76.40\pm1.21$            & $76.20\pm1.11	$            & $\mathbf{79.49	\pm0.98}$ \\
         \bottomrule
    \end{tabular}
    }
    \label{acoustic_comp}
\end{table}

\subsection{Prosody Models}
As before, we begin by finding the best temperature $\tau$ for the Flow- and Diffusion-based prosody models. 
We keep the acoustic model (\textit{Flow-AM}) fixed and %
condition this with the generated \emph{f0} and duration from the various prosody models. 
We evaluate the speech samples in terms of naturalness, with results shown in \autoref{prosody_temps_comp}. Following the same criterion, we choose $\tau=0.4$ and $\tau=0.8$ for \textit{Flow-PM} and \textit{Diff-PM}, respectively.

\begin{table}[htbp]
        \caption{MUSHRA naturalness evaluation results for temperature $\tau$, showing mean values with 95\% confidence intervals. * indicates %
        no statistically significant difference from the highest MUSHRA score.}
    \centering
    \scalebox{0.7}{
    \begin{tabular}{m{1.5cm} m{1.7cm}m{ 1.7cm}  m{1.7cm}  m{1.7cm}}
         \toprule 
         System        &  $\tau=0.2$                       &  $\tau=0.4$  & $\tau=0.6$  & $\tau=0.8$ \\
         \midrule
         
         \textit{Flow-PM}   & $\mathbf{74.36\pm1.10}$  & ${73.94\pm 1.12}^{*}$ & $72.28\pm1.10	$ & $70.74\pm1.10$ \\
         \textit{Diff-PM}   & $\mathbf{78.40\pm 0.83}$          & ${78.27\pm0.84}^{*}$ & $ {77.92\pm0.87}^{*}$  & ${77.71\pm0.88}^{*}$  \\
         
         \bottomrule
    \end{tabular}
    }
    \label{prosody_temps_comp}
\end{table}

\begin{figure}[t]
  \centering
  \includegraphics[width=0.55\linewidth]{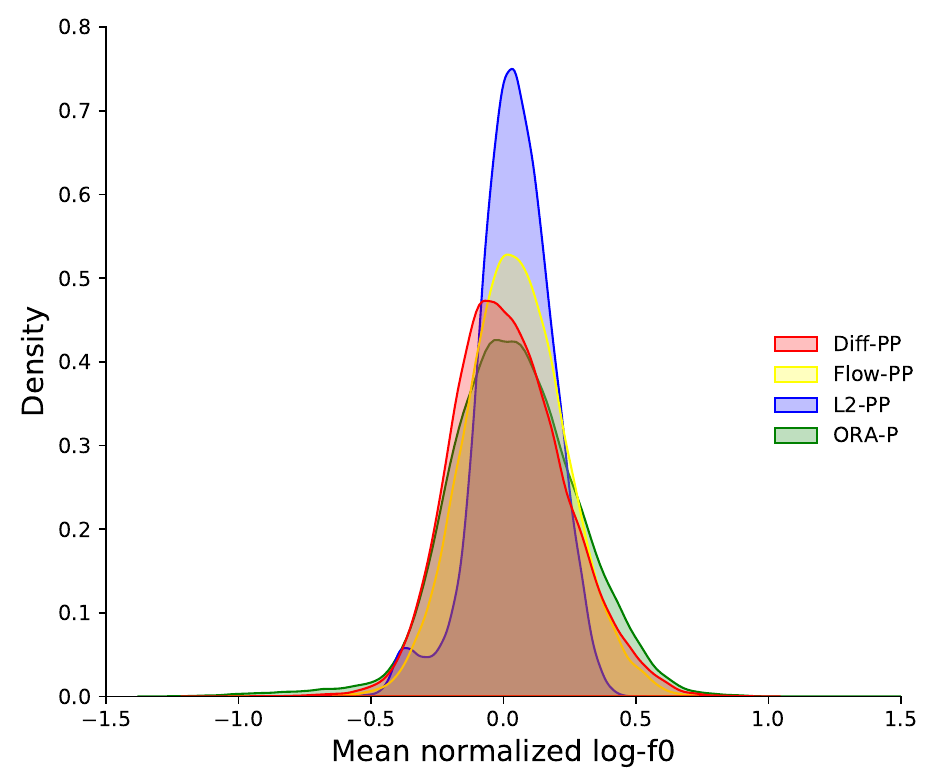}
  \caption{Distribution of \emph{log-f0} values generated by the different Prosody Models.}
  \label{fig:histogram of generated prosody}
\end{figure}

We present the results for objective metrics in \autoref{table:obj_eval_stats}.
Results %
find that 
\textit{Flow-PM} and \textit{Diff-PM} %
produce features with higher standard deviations than \textit{$\mathcal{L}2$-PM}. %
This confirms 
that the proposed prosody models are able to mitigate the over-smoothing %
problem and produce more dynamic prosody features, which %
can lead 
to more expressive speech. 
Considering JSD, we observe that \textit{Diff-PM} and \textit{Flow-PM} %
produce features that are closer to the distribution of oracle features. 
This can also be seen by the \emph{log-f0} distribution from the different prosody models in \autoref{fig:histogram of generated prosody}.
Most of the \emph{f0} values from \textit{$\mathcal{L}2$-PM} are concentrated around 0.
In contrast, the distributions of \emph{f0} from the other two systems are more dispersed and have longer tails, indicating better distribution coverage.

\begin{table}[htbp]
    \caption{Standard deviation (STD) and Jensen-Shannon divergence (JSD) for the different prosody models.}
    \centering
    \scalebox{0.7}{
    \begin{tabular}{llllll}
        \toprule 
        \multirow{2}{*}{System} & \multicolumn{3}{c}{STD} & %
        \multicolumn{2}{c}{JSD} \\
        \cmidrule(lr){2-4} %
        \cmidrule(lr){5-6}
                                & \emph{log-f0}        & dur       & $\Delta$\emph{log-f0}   & \emph{log-f0}          & dur        \\
                                \midrule
        \textit{ORA-P}         & 0.24      & 4.49      & 0.19 & -          & - \\
        \midrule
        \textit{$\mathcal{L}2$-PM}         & 0.14      & 4.03      & 0.09 & 0.163      & 0.112 \\
        \textit{Flow-PM}        & 0.19      & 4.12      & 0.14 & 0.073      & 0.067 \\
        \textit{Diff-PM}        & $\mathbf{0.21}$      & $\mathbf{4.27}$      & $\mathbf{0.15}$ & $\mathbf{0.057}$      & $\mathbf{0.041}$ \\
        
        \bottomrule
        \end{tabular}
        }
        \label{table:obj_eval_stats}
\end{table}

\autoref{prosody_predictor_comp} shows the results %
from the subjective evaluations of the prosody models.
There are no significant differences between the three competing systems and \textit{ORA-P} in terms of naturalness. 
This finding is perhaps unexpected and suggests that there is little room for naturalness improvement, on average across all speech samples.
However, when we consider only utterances from speaking styles with high arousal, such as anger or happiness \cite{kensinger2006processing}, %
\textit{Diff-PM} and \textit{Flow-PM} have a larger gap to \textit{$\mathcal{L}2$-PM}, with \textit{Flow-PM} outperforming \textit{$\mathcal{L}2$-PM}.
The results indicate that \textit{Flow-PM} and \textit{Diff-PM} can contribute most to prosody modelling for styles with high arousal.
In terms of style similarity and expressiveness, no %
significant differences are found between %
\textit{Flow-PM}, \textit{Diff-PM} and \textit{ORA-P}.
Both \textit{Flow-PM} and \textit{Diff-PM} %
are found to significantly outperform \textit{$\mathcal{L}2$-PM}. 
It is somewhat surprising that \textit{Diff-PM} and \textit{Flow-PM} are both on par with \textit{ORA-P} in terms of naturalness, style similarity and expressiveness.
Specifically, %
the objective analysis in \autoref{table:obj_eval_stats} found the oracle prosody features to have %
larger standard deviations than %
those from \textit{Diff-PM}  and \textit{Flow-PM}, %
however it appears as though these differences do not result in listener preferences. 
A possible explanation for this could be that the expressiveness in speech also depends on how the acoustic model %
represents the prosody features. 

We conducted follow-up preference tests for \textit{Diff-PM} and \textit{Flow-PM} in terms of naturalness, style similarity and expressiveness.
However, %
no significant differences were found between \textit{Flow-PM} and \textit{Diff-PM} for any of the metrics.
Overall, \textit{Flow-PM} and \textit{Diff-PM} are on par with each other, but significantly preferred to \textit{$\mathcal{L}2$-PM}.

\begin{table}[htbp]
        \caption{Mean MUSHRA scores for prosody models, along with 95\% confidence intervals.}
    \centering
    \scalebox{0.8}{
    \begin{tabular}{cccc}
         \toprule 
         Method             &  Naturalness              & Style Similarity          & Expressiveness \\
         \midrule
         \textit{ORA-P}       & $79.61\pm0.90$             & $78.62\pm0.91$            & $77.20\pm0.92	$ \\
         \midrule
         \textit{$\mathcal{L}2$-PP}    & $79.02 \pm0.94$            & $76.49\pm0.98	$            & $75.74\pm0.99	$ \\
         \textit{Flow-PM}   & $79.41	\pm0.88		$   & $\mathbf{78.23	\pm0.91	}$   & $77.17	\pm0.89	$ \\
         \textit{Diff-PM}   & $\mathbf{79.51\pm0.92}$            & $78.10 \pm0.92$            & $\mathbf{77.32	\pm 0.91}$ \\
         \bottomrule
    \end{tabular}
    }
    \label{prosody_predictor_comp}
\end{table}

\section{Discussion}
Both flow and diffusion approaches learn a mapping of the target features, coming from complex unknown distributions, transforming them to points %
from %
defined simple prior distributions. Losses applied during training are made relative to the likelihood of the prior distribution.
However, $\mathcal{L}$1/$\mathcal{L}$2-based models place a strong assumption that the distribution of the target features is a Gaussian and return values that come from the mean of the distribution, resulting in less expressive predictions.
We hypothesise that the U-Net structure within the diffusion model is good at %
generating features considering long-term dependencies, %
a desirable trait for prosody modelling. However, %
for %
acoustic modelling, when provided with prosody conditioning the long-term information is already being largely explained to the model. Instead, %
the acoustic model is being asked to focus on short-term quality of generated individual spectrogram frames. %
For such a task it appears that the diffusion model does not perform as well as flow or $\mathcal{L}$1-based models. 

\section{Conclusion}
In this paper, we study and compare three different methodologies for acoustic and prosody modelling: normalizing flows, diffusion probabilistic models, and models trained with $\mathcal{L}$1/$\mathcal{L}$2 loss. For acoustic modelling, subjective evaluation results suggest that an acoustic model based on Normalizing Flows achieves the best results. %
For prosody modelling, we observe comparable performance for flow-based and a diffusion-based models in terms of naturalness, style similarity and expressiveness. In terms of both objective and subjective evaluation, the prosody features predicted from flow-based and diffusion-based models demonstrate %
improved expressiveness and better style similarity than the prosody model optimized using an $\mathcal{L}$2 loss.

\bibliographystyle{IEEEtran}
\bibliography{references}

\end{document}